\documentclass[a4paper,10pt, twocolumn]{article}
\usepackage{graphicx}
\usepackage{amsmath}
\usepackage{amssymb}
\usepackage{authblk}
\usepackage[colorlinks=true, allcolors=blue]{hyperref}
\usepackage{fancyhdr}
\usepackage[titletoc]{appendix}
\usepackage{multicol}
\usepackage{subcaption}
\usepackage{bm}
\usepackage{bm}
\usepackage{amsfonts}
\usepackage{epstopdf}
\usepackage{xcolor}
\usepackage{color}
\usepackage{wasysym}
\usepackage[english]{isodate}
\usepackage[switch,columnwise]{lineno} 
\usepackage{ulem}
\usepackage[capitalise]{cleveref}
\usepackage{authblk}
\usepackage{upgreek}

\usepackage{makecell}
\usepackage{hhline}
\usepackage{setspace}
\usepackage{etoolbox}

\usepackage{afterpage}
\newcommand\redsout{\bgroup\markoverwith{\textcolor{red}{\rule[0.5ex]{2pt}{0.4pt}}}\ULon}
\makeatletter
\patchcmd{\@maketitle}{\LARGE}{\large}{}{}

\makeatother
\newcommand{\micron}{{\upmu\mathrm{m}}}
\newcommand{\Wcmsqd}{{\mathrm{W }\mathrm{cm}^{-2}}}

 \usepackage{authblk}

\title{Efficient generation of axial magnetic field  \\ by multiple laser beams with twisted pointing directions}% Force line breaks with \\
\author[1]{Yin Shi}
\author[2]{Alexey Arefiev}
\author[1]{Jue Xuan Hao}
\author[1,3]{Jian Zheng}

\affil[1]{Department of Plasma Physics and Fusion Engineering,\authorcr  University of Science and Technology of China, Hefei 230026, China}
\affil[2]{Department of Mechanical and Aerospace Engineering, \authorcr 
University of California at San Diego, La Jolla, CA 92093, USA}
\affil[3]{Collaborative Innovation Center of IFSA,\authorcr 
Shanghai Jiao Tong University, Shanghai 200240, Peoples Republic of China}%Lines break automatically or can be forced with \\

\date{\today}% It is always \today, today,
\begin{document}

%\doublespacing
%\onehalfspacing
\twocolumn[\begin{@twocolumnfalse}
\maketitle

\begin{abstract}
    Strong laser-driven magnetic fields are crucial for high-energy-density physics  and laboratory astrophysics research, but generation of axial multi-kT fields remains a challenge. The difficulty comes from the inability of a conventional linearly polarized laser beam to induce the required azimuthal current or, equivalently, angular momentum (AM). We show that several laser beams can overcome this difficulty. Our three-dimensional kinetic simulations demonstrate that a twist in their pointing directions {enables them to carry orbital AM and transfer it to the plasma, thus generating a hot electron population carrying AM needed to sustain the magnetic field.} The resulting multi-kT field occupies a volume that is tens of thousands of cubic microns and it persists on a ps time scale. The mechanism can be realized for a wide range of laser intensities and pulse durations.  Our scheme is well-suited for implementation using {multi-kJ PW-class lasers, because, by design, they have multiple beamlets and because the scheme requires only linear-polarization.}
\end{abstract}
\vspace{5mm}
%\vfill
 \end{@twocolumnfalse} ]
%\maketitle

%\linenumbers
%\tableofcontents

%\section{\sffamily{Introduction}}
Recently, magnetic field effects in high energy density physics (HEDP) have attracted significant interest~\cite{Moody2022a, lmj_bfield2022, Santos2018pop}. {These can range from guiding of relativistic electron beams~\cite{Bailly-Grandvaux2018} to affecting the shape of inertial fusion implosions~\cite{Bose2022}.} Despite significant progress,  generation of sufficiently strong and controllable macroscopic fields at the laser facilities used for HEDP research~\cite{Danson2019, li2022, zhu2018hpl,Kawanaka_2008,Crane_2010,Batani_2014} remains an outstanding challenge. 

{Various approaches to magnetic field generation using high-power lasers have been explored in search of an optimal mechanism and field configuration. Initial efforts were focused on leveraging a circularly polarized (CP) laser beam to generate an axial quasi-static plasma magnetic field~\cite{ Haines2001, Sheng1996, Najmudin2001}. The emergence of capabilities to create Laguerre-Gaussian (LG) high-intensity beams has stimulated research into generation of the axial field using such beams as well~\cite{Ali2010, Shi2018, Vieira2018, Nuter2018}. The strength of the plasma field is limited by the laser's ability to drive a strong azimuthal current, so it is insightful to interpret the process as a transfer of the laser's angular momentum $\bm{L}$ to the plasma. Here $\bm{L} = \varepsilon_0 \int \bm{r} \times \left[ \bm{E} \times \bm{B} \right] d^3 \bm{r}$, where $\varepsilon_0$ is the dielectric permittivity, $\bm{E}$ and $\bm{B}$ are the electric and magnetic fields, respectively.}

{Setups involving conventional linearly polarized (LP) laser beams have also received attention, because additional optics is required to make CP or LG beams from the conventional beams. A large-scale uniform magnetic fields can be created by a ns laser irradiating a capacitor–coil~\cite{Santos2018pop, Sakata2018, Kochetkov2022, Morita2023} or a snail target~\cite{Pisarczyk2023}. This field can then be amplified inside a plasma by a high intensity ps or sub-ps laser pulse ~\cite{Murakami2020, Shi2020, Wilson_2021,Zosa2022}. Relativistic electrons generated by high-intensity laser pulses  can also generate surface or bulk azimuthal magnetic fields when streaming through a solid density target~\cite{Davies1999, Robinson2014, huang2019, cai2020}, and these fields 
are beneficial for hot electron transport and electron beam collimation~\cite{Sheng1996, Najmudin2001, Gorbunov1996, Sheng1998, Bell2003, zheng2004, Kaymak2016, ZhangF2020}.
Applications for longitudinal fields include guiding of relativistic electron beams~\cite{wangwm2015, Santos2018pop,Sakata2018}, laser-driven ion acceleration~\cite{ Arefiev2016, Weichman2022pop}, magnetized atomic physics~\cite{Lai2001rmp, Santos2018pop}, and laboratory astrophysics~\cite{wang_Li2016}. }

\begin{figure}[!ht]
\centering
\includegraphics[width=0.97\columnwidth]{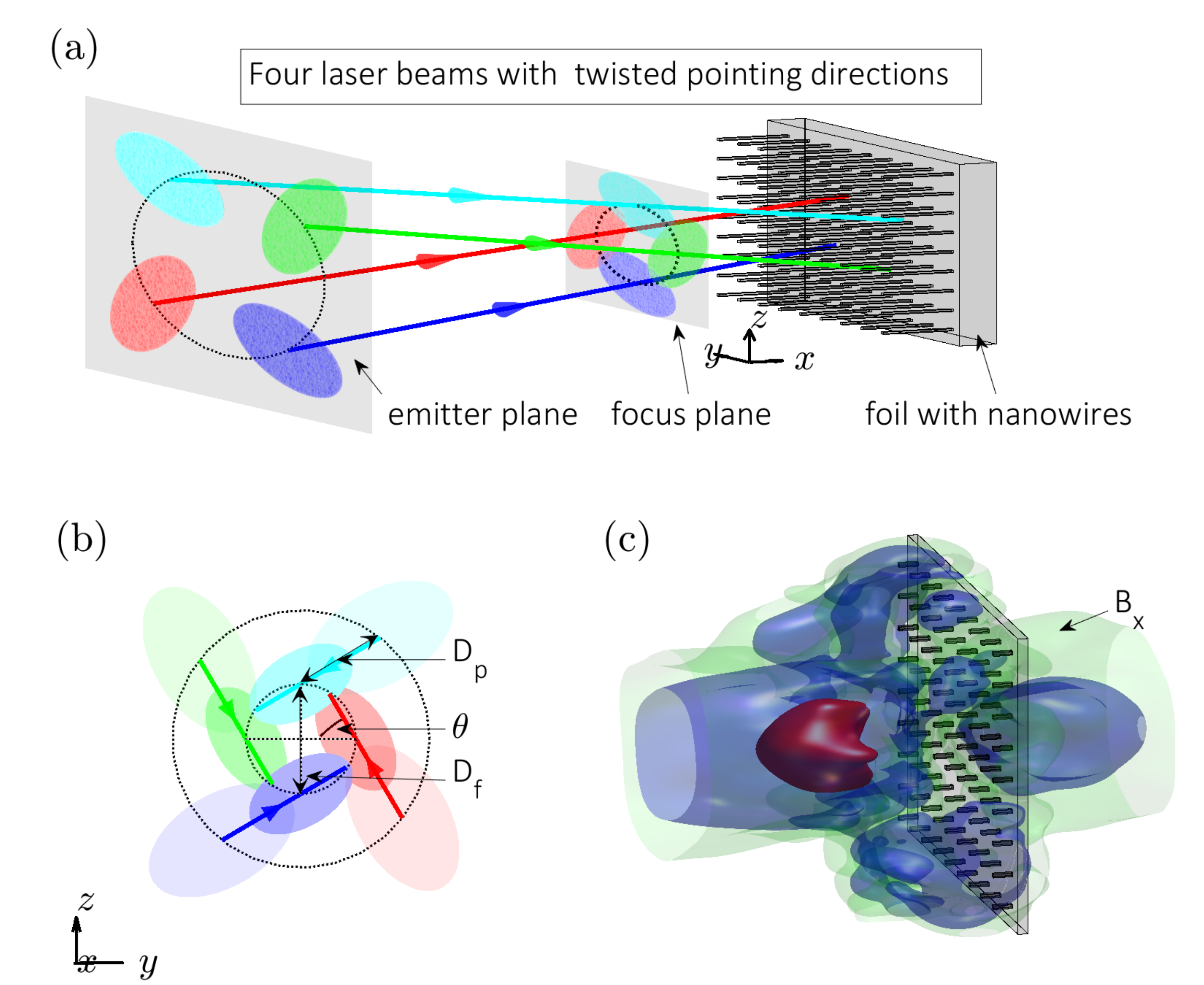}
\caption{(a)~Setup for axial magnetic field generation using four {linearly-polarized Gaussian} laser beams with twisted pointing directions, shown with solid lines, and a structured target. The size of each beam is shown with a color-coded ellipse in the emitter plane (left side of the simulation box) and in the focus plane. (b) Projections of the two planes on to the $(y,z)$-plane. The parameters setting up the beam orientation are defined in the text. (c) Surface plots of the axial magnetic field $B_x$ after the lasers have left the simulation box ($t = 20$~fs). The  green, blue, and red surfaces represent $B_x/B_0 = -0.1$, $-0.2$, and $-0.8$, where $B_0 = 13.4$~kT. } \label{Scheme}
\end{figure} 

{Generation of a large-volume strong magnetic field requires significant energy that must be delivered by the laser. Multi-kJ PW-class laser systems like LFEX~\cite{Kawanaka_2008}, NIF ARC~\cite{Crane_2010}, and Petal~\cite{Batani_2014} offer the highest energy that can be delivered on a ps time scale. These lasers are all composed of multiple LP beamlets. The multi-beamlet configuration is not just an essential feature of the laser system design, but  also the key to advanced laser-plasma interaction regimes~\cite{Morace2019}. The number of multi-beamlet facilities will increase, as SG-II UP~\cite{zhu2018hpl} is due to be upgraded to have multiple kJ-class ps laser beams.}

{This Letter presents a new multi-beam approach for efficient laser-to-plasma angular momentum (AM) transfer resulting in magnetic field generation. The approach, illustrated in \cref{Scheme} for four linearly-polarized Gaussian beams, is motivated by the capability of multi-kJ PW-class laser systems to provide multiple beamlets~\cite{Kawanaka_2008, Crane_2010, Hernandez-Gomez_2010, Batani_2014}.} Our scheme eliminates the need for CP or LG beams while offering a method for generating a field above $10$~kT in a $10^4~\micron^{3}$ volume.   Our scheme provides a plasma that can potentially be used for studies of astrophysical objects involving strong magnetic fields beyond the dynamic range of previous laboratory settings~\cite{Beloborodov2022, takabe_kuramitsu_2021} and to mimic a rotating plasma environment in astrophysics~\cite{Tamburini2011, Tamburini2019}. 

{The role of the twist in the pointing direction, \cref{Scheme}(a), can be illustrated using geometrical optics. Each laser beam is represented by a ray directed along the wave vector $\bm{k}_i$,} where $i$ is the index numbering the beam. The photon momentum in the $i$-th beam is $\bm{p}_{i} = \hbar \bm{k}_{i}$. {Consider a pair of tilted rays, $\bm{k}_{1,2} = (k_x, k_{\perp}^{(1,2)}, 0)$, that intersect the $(y,z)$-plane at $z_{1,2} = \pm D_f/2$ and $y_{1,2} = 0$, where $D_f/2$ is the beam offset. The axial AM of a photon is $[\bm{r} \times \bm{p}]_x$, so the total AM of the two beams is $L_x \approx - N \hbar ( k_{\perp}^1 - k_{\perp}^2) D_f/2$, where $N$ is the number of photons in each beam. 
%If the two beams have the same tilt, then $k^{(1)}_{\perp} = k^{(2)}_{\perp}$ and $L_x \approx 0$. If the tilt of the second beam is opposite to that of the first, $k^{(2)}_{\perp} = -k^{(1)}_{\perp}$, then $L_x \approx - N \hbar k_{\perp}^{(1)} D_f$. 
The AM can be doubled by adding two rays offset in $y$. The rays appear twisted, so it is appropriate to refer to the calculated AM as orbital angular momentum (OAM). They carry OAM, a distinct form of AM, even though each beam has no intrinsic AM~\cite{Bliokh2015}.} There are parallels to $\gamma$-ray beams carrying OAM~\cite{liu2016, chen2018, chen2019} composed of photons with a twisted distribution of $\bm{p}$. 

%+++++++++++++++++++++++++++++++++++++++++++++++++++++++++++++++++++++++++++++++++++++++++++++++++++++++++++

To investigate the transfer of the OAM carried by four laser beams, we have performed a series of three-dimensional (3D) particle-in-cell (PIC) simulations using a relativistic PIC code EPOCH~\cite{arber2015contemporary}. Each beam is a linearly polarized Gaussian beam. 
The duration of each pulse is 450~fs and the peak intensity is $2.1 \times 10^{20}~\Wcmsqd$. Our target is a flat foil with sub-wavelength diameter nanowires whose purpose is to increases the interaction volume between the laser beams and the plasma produced by the target and thus enhances the number of hot electrons~\cite{ji2016prl}. The front and rear surfaces of the foil are located at $x_{f1} = 0~\micron$ and $x_{f2} = 4~\micron$. The spacing between the wires is $2~\micron$, the wire length is $5~\micron$, and the wire width is $0.4~\micron$. The entire target is initialized as {a fully ionized cold carbon plasma } with an electron density of $50n_c$, where $n_{c} = 1.8\times 10^{21}$~cm$^{-3}$ is the critical density corresponding to a laser wavelength $\lambda = 0.8~\micron$. {All simulation parameters are listed in the Supplemental Material.}

The orientation of the four beams is set according to \cref{Scheme}. Their axes intersect a given plane perpendicular to the $x$-axis with the intersection points forming vertices of a square. We use two planes: the emitter plane ($x_e = - 20~\micron$), which is the left boundary of the simulation box, and the focus plane ($x_f = - 16~\micron$), which is the plane where the beams have the smallest transverse size. The twist is set by angle $\theta$. There is no twist for $\theta = 0$, so that the axis of each beam and the $x$-axis form a plane. We use $\phi = \arctan(-D_p/S)$ to set the beam convergence, where $S$ is the distance between the emitter plane and the focus plane and $D_p$ is the transverse shift of the beam axes between the two planes, as shown in \cref{Scheme}(b). We use $\phi = - 0.27 \pi$ in all simulations.

\Cref{Scheme}(c) shows the magnetic field (B-field) for $\theta= - 0.28 \pi$ at $t = 20$~fs. We define $t = 0$~fs as the time when the laser pulses leave the simulation box. {The laser-plasma interaction takes place at $t \in (-510, -60)$~fs.} The longitudinal B-field exceeds 10~kT. The volume is around $10^4~\micron^3$. The three surfaces show $B_x/B_0 = -0.1$, $-0.2$, and $-0.8$, where $B_0 = 13.4$~kT. Note that $B_0 \equiv 2 \pi m_e c/|e| \lambda_L$, where $\lambda_L = 0.8~\micron$ is laser wavelength in vacuum, $c$ is the speed of light, and $e$ and $m_e$ are the electron charge and mass. To make its profile more clear, $B_x$ is averaged temporally over a 20~fs interval and spatially using a box with stencil size $0.4~\micron \times 0.4~\micron \times 0.4~\micron$.

\begin{figure}[!ht]
    \centering
    \includegraphics[width=0.97\columnwidth]{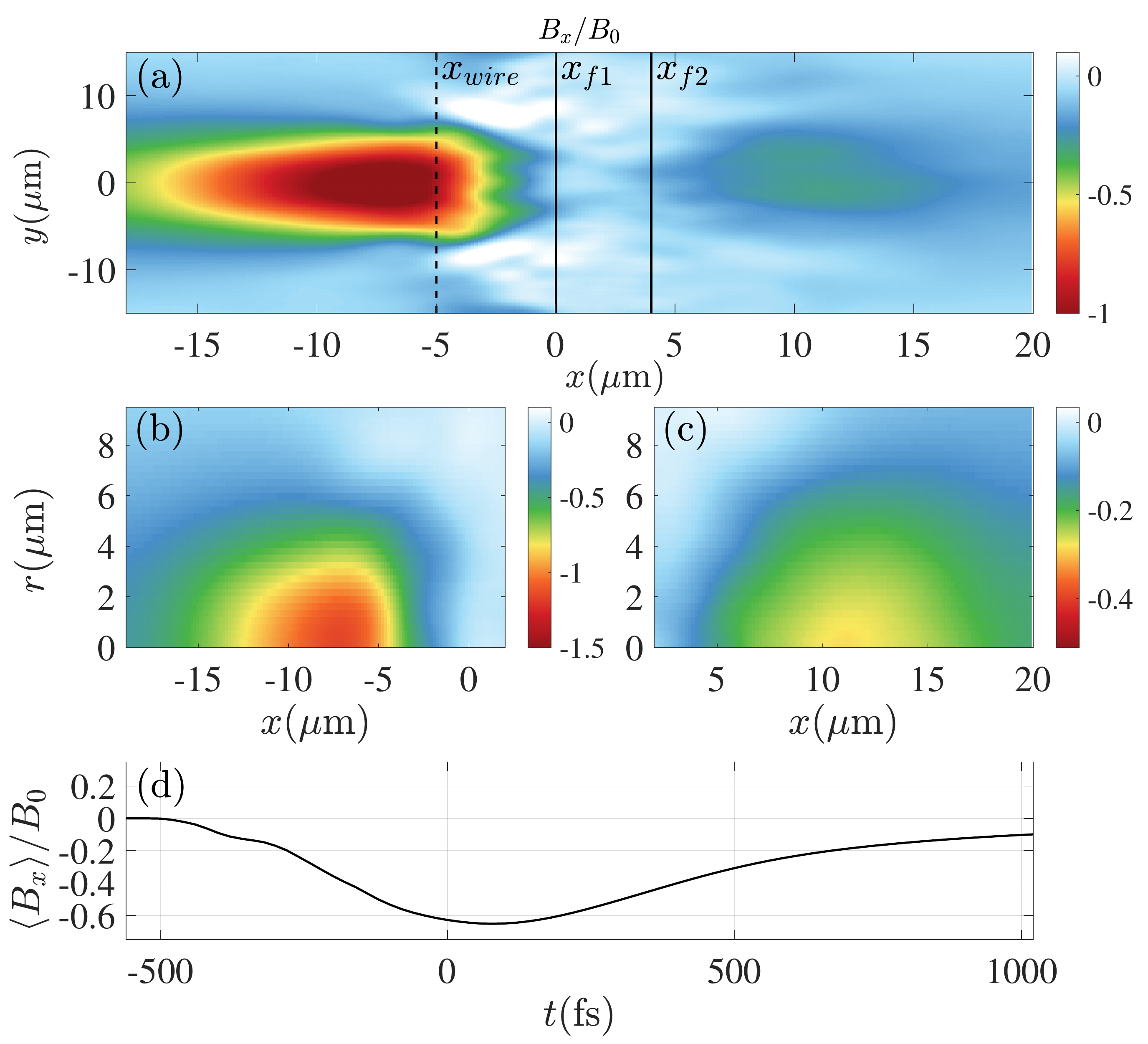}
\caption{(a) Axial magnetic field in the $(x, y)$-plane at $t = 20$~fs. (b)\&(c) Angle-averaged axial magnetic field as a function of $x$ and $r$ at $t = 20$~fs. The nanowire region is at $x_{wire} < x < x_{f1}$. The foil is at $x_{f1} \leq x \leq x_{f2}$. (d) Time evolution of the volume-averaged magnetic field within a box with $-15~\micron<x < x_{wire}$, $|y| < 5~\micron$, and $|z| < 5~\micron$.} \label{Bx}
\end{figure}

 Owing to the approximately axisymmetric profile of $B_x$, we can examine its 2D distributions in \cref{Bx} without missing too much information. \Cref{Bx}(a) shows the global distribution in the $(x, y)$-plane. The nanowires are between $x_{wire} = -5~\micron$ and $x_{f1} = 0~\micron$. The foil is between $x_{f1} = 0~\micron$ and $x_{f2} = 4~\micron$. Figures~\ref{Bx}(b)\&(c) show $B_x$, averaged over the azimuthal angle, as a function of $x$ and $r$ in front of and behind the target (note the different color-scale ranges).  We find that $|B_x|$ can be as high as $1.5 B_0$ in front of the wires because the lasers generate a higher concentration of hot electrons in front of the foil. Reaching this amplitude is noteworthy because new phenomena of laser beam transport through a plasma can arise at $|B_x| \gtrsim B_0$~\cite{wudong2020, KLi2022}. Even though $B_x$ is weaker behind the target, it is in the range of kT. This confirms that our scheme indeed produces electrons carrying AM, as the lasers are unable to reach behind the target to generate the B-field locally.

\Cref{Bx}(d) shows the time evolution of the average magnetic field strength $\langle B_x \rangle$ in a box with  $-15~\micron<x <-5~\micron$, $|y| < 5~\micron$, and $|z| < 5~\micron $. 
The ps time scale is comparable to that in Ref.~\cite{Longman2021}, but the region containing the magnetic field moves axially outward (away from the target).
In terms of the energy content within a region with  $|y|$, $|z| < 15~\micron$, we find that the energy in the magnetic field ($\varepsilon_B = \int B_{x}^{2}/(2 \mu_0) dV \approx 3.0$~J) is much smaller than the kinetic energy of electrons ($\varepsilon_{e}\approx 40.0$~J). The energy of the four beams is $\varepsilon_{laser} \approx 580$~J. The energy conversion efficiency from laser to hot electrons and from hot electrons to the magnetic field are both around 10\%. The overall conversion efficiency is two orders of magnitude higher than that for a laser-driven coil in Ref.~\cite{lmj_bfield2022}.

\begin{figure}
\centering
\includegraphics[width=0.97\columnwidth]{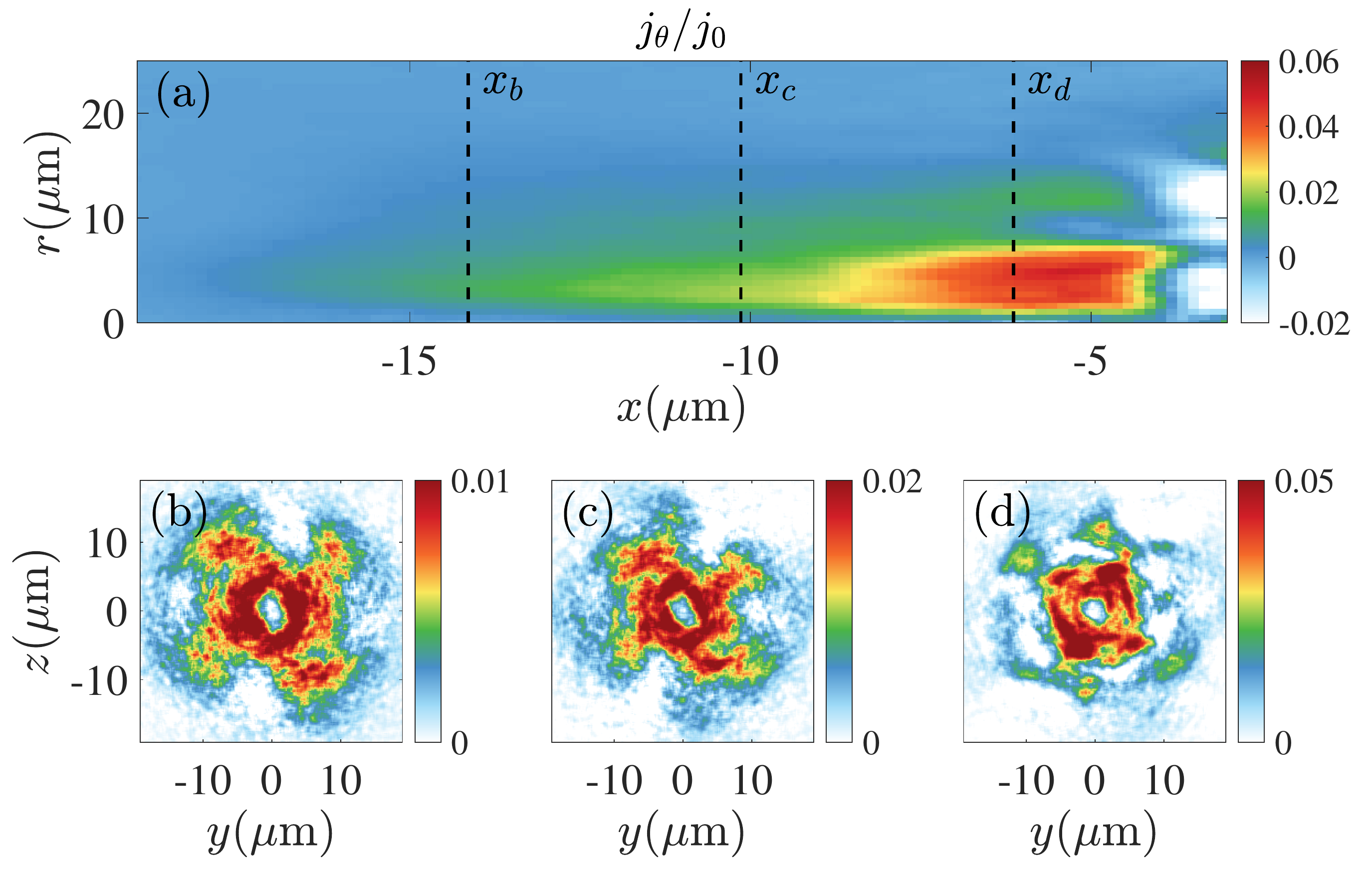}
\caption{(a) Angle-averaged azimuthal current density $j_{\theta}$ at $t = 20$~fs as a function of $x$ and $r$. (b), (c), and (d) $j_{\theta}$ in the $(y,z)$-plane at three different locations with $x = x_b$, $x_c$, and $x_d$. Note the significant difference in color-scales between the three panels introduced to improve visibility. The current density is normalized to  $j_0 = -|e| c n_c = - 8.25\times 10^{16}$~A/m$^2$.}\label{jtheta}
\end{figure}

To confirm that the twist angle $\theta$ {rather than the polarization} is the key parameter, we performed a simulation without the twist ($\theta = 0$) and a simulation with an opposite twist to the original direction ($\theta= 0.28 \pi$). We found that no axial magnetic field is generated without the twist and that $B_x$ reverses its direction when we reverse the twist. The azimuthal B-field is generated in all three cases due to the ubiquitous axial current driven by the laser pulses. {We also performed a simulation with the original setup but randomly selected direction of the E-field polarization in each laser beam. The angle-averaged $B_x$ is similar to the $B_x$ in \cref{Bx}, confirming that laser polarization has only a secondary effect on the magnetic field generation in our setup.} 

In the remainder of this letter we focus on the region in front of the target. We start with an analysis of the azimuthal current density $j_\theta$ that is thought to be responsible for the axial magnetic field generation. \Cref{jtheta}(a) shows $j_\theta$, averaged over the azimuthal angle, in the $(x, r)$-plane at $t = 20$~fs. The direction of $j_\theta$ alternates in the nanowire region [$x \in (-5, 0)~\micron$]. The underlying cause is the presence of strong nonuniformities in the ion density associated with the original nanowires.

Transverse distributions of $j_\theta$ in the $(y, z)$-plane at different $x$ positions ($x_b$, $x_c$, $x_d$) are shown in \cref{jtheta}(b-d). These positions are marked by dashed lines in \cref{jtheta}(a). In agreement with \cref{jtheta}(a), $|j_\theta|$ is the biggest at $x_d$ and the smallest at $x_b$. To perform an order of magnitude estimation for the maximum value of $|B_x|$, we assume that $j_\theta$ is uniform inside a cylinder of radius $R$ and length $\Delta x$. Then the Biot-Savart law~\cite{jackson1975electrodynamics} yields
\begin{eqnarray}\label{bx_cal}
    \max |B_x| & \approx & \dfrac{\mu_0}{2} \int_0^R \int_{-\Delta x}^{\Delta x}  \dfrac{|j_{\theta}| r^2 \mathrm{d}x \mathrm{d}r}{(r^2 + x^2)^{3/2}}    \nonumber \\
    &= &\mu_0 |j_{\theta}| \Delta x \operatorname{arsinh}(R/\Delta x),
\end{eqnarray}
where $\mu_0 = 1.26 \times 10^{-6}$~H/m is permeability in vacuum. According to \cref{jtheta}(a), we can set $R \approx \Delta x \approx 5~\micron$. In \cref{jtheta}(d), the current density reaches $|j_{\theta}| \approx 0.05 |j_0|$, where $j_0 \equiv -|e| c n_c $.  Using this value, we obtain $\max |B_x| \sim 20$~kT, which is close to the peak magnetic field, $ B_{x} \sim 1.5 B_0$, in \cref{Bx}(b). 

To quantify the rotating effect of the plasma, we computed the density of the axial AM for electrons and ions. Due to the significant difference in mass, the ratio of the axial AM absorption between electrons and ions is $\eta_{ei} \approx 0.01$. We can estimate the AM density of hot electrons using the azimuthal current density. We write the AM density of electrons as $L_{xe} \approx r \gamma_{a} m_e n_e v_{\theta}$, where $\gamma_{a}$ is the relativistic gamma-factor, $n_e$ is the number density, and $v_{\theta}$ is the effective azimuthal velocity. 
We set $v_{\theta} \approx - j_{\theta}/|e|n_e$ to find that $L_{xe} \approx r \gamma_{a} m_e j_{\theta}/|e|$. For $r = w_0$ and $\gamma_{a} \approx \sqrt{1 + a_0^2} \approx 10$, we have $L_{xe} \approx 1.6$~kg/m-s.
Using the electron density from simulations, $n_e \approx 10^{27}$~m$^{-3}$, we find that the rotating velocity is around $v_\theta \approx 0.1c$. Our setup produces a rotating plasma environment with electron density and rotation velocity two orders of magnitude higher than an LG beam in~\cite{Shi2018}.

The OAM transfer from the laser beams to the electrons can be determined using the conservation of AM~\cite{Haines2001, Ali2010, Longman2021}. The OAM of absorbed laser photons is transferred to electrons and ions, with the electron fraction equal to $\eta_{ei}$. Then, based on the photon absorption, the axial AM density of the electrons is roughly
\begin{eqnarray}  \label{oam_abs}
   L_{ex}(x, r) &\approx&  \frac{\eta \eta_{ei}}{x_{\kappa}} \frac{0.75\tau_g I_0}{c} \sin(\phi) \sin(\theta)D_{xr}, \nonumber \\
    D_{xr}&=& r e^{x/x_{\kappa}- 2(r - D_f/2)^2/w_0^2},
\end{eqnarray}
where $I_0$ is the peak intensity of the incident laser pulses and $\tau_g$ is their duration. For simplicity, we assume that the absorption coefficient $f_{abs}$ of the laser intensity over the axial distance is $f_{abs} = f_{0} \exp(x/ x_{\kappa})$. We find from the simulation that $\eta = \int_{x = x_e}^{x = 0} f_{abs} dx \approx x_{\kappa} f_0 \approx 0.1$ ($x_{\kappa} \approx 3 ~\micron$). We use $r = D_f/2 = 6~\micron$ and $x = 0~\micron$ to find the peak AM density of the electrons, $L_{ex} \approx 2.4$~kg/m-s. {This result is on the same scale as the peak AM density ($\approx$ 6.9~kg/m-s) in our simulations. It is also close to the result ($\approx$ 1.6~kg/m-s) calculated using $j_{\theta}$ in \cref{jtheta}. The peak AM density in the simulation exceeds our model's prediction, which may be due to the locally positive AM density in the nanowire region.}
According to Eq.~(\ref{oam_abs}), the axial B-field can be controlled by changing the sign of twist angle $\theta$, which has been confirmed in the Supplemental Material. Our model ignores the dependence of $f_{abs}$ on such parameters like $I_0$, $\phi$, and $\theta$, but the actual {absorption mechanism may be more complex~\cite{Brunel1988, Levy2014, Grassi2017}}. Using $L_{ex}$, we can obtain the azimuthal current density and the associated axial B-field, 
\begin{eqnarray}  \label{bx_theory}
  B_{x}  \propto  \frac{j_{\theta}(x, r)}{j_0}  \propto  (\eta \eta_{ei})\frac{a_0^2}{\gamma_{a}}\frac{c\tau_g}{x_{\kappa}} \frac{D_{xr}}{r} \sin(\phi) \sin(\theta). 
 \end{eqnarray}

\begin{figure}[tbp!]
    \centering
    \includegraphics[width=0.97\columnwidth]{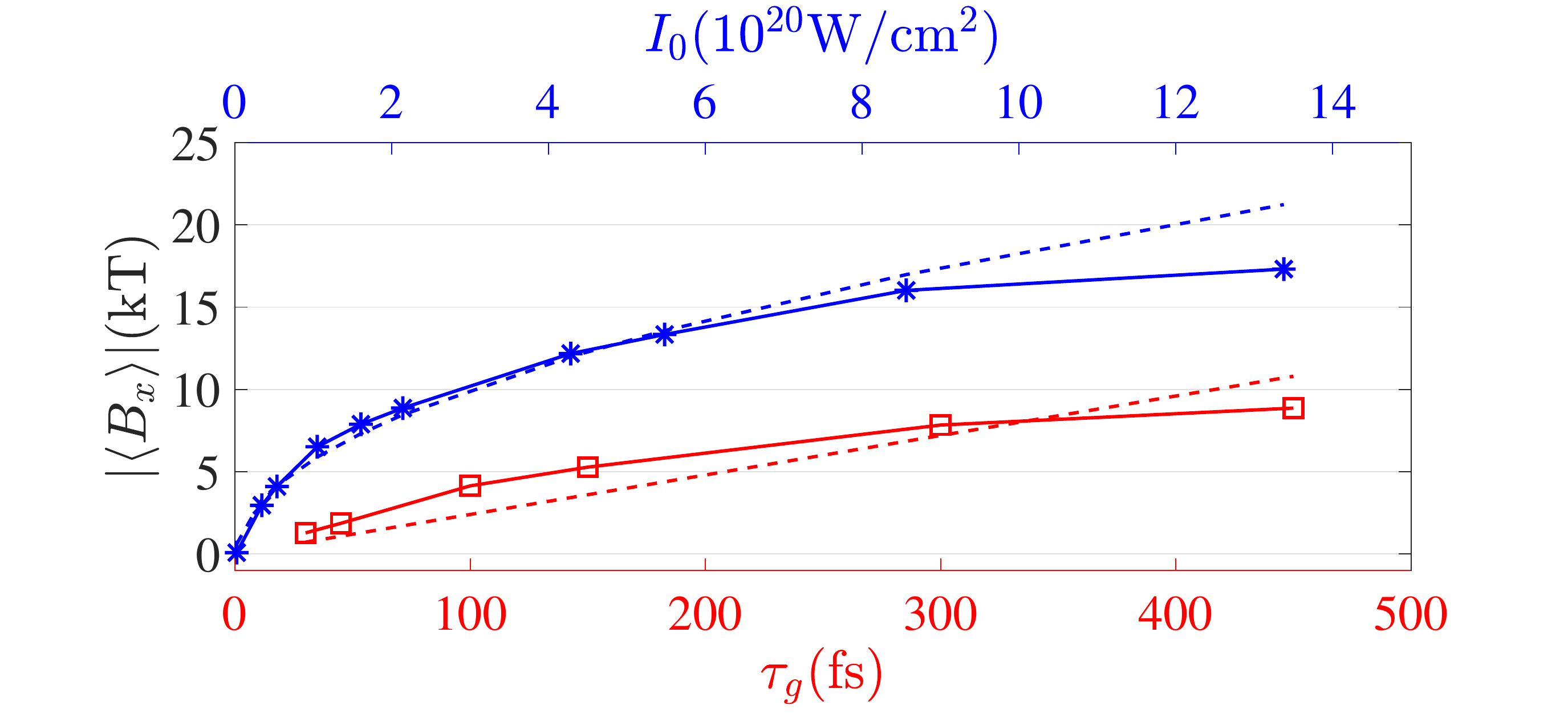}
\caption{Volume-averaged magnetic field $|\langle B_x \rangle|$ as a function of peak laser intensity $I_0$ (blue asterisk markers) and laser pulse duration $\tau_g$ (red square markers). The averaging is performed within a box with $-15~\micron<x < x_{wire}$, $|y| < 5~\micron$, and $|z| < 5~\micron$. The dashed curves (blue and red) show the fits based on Eq.~(\ref{bx_theory}).
} \label{bxav_scan}
\end{figure}
% $L_0 = cm_e n_c w_0 = 3e8*9.1e-31*1.8e27*7.0e-6 = 3.44 kg/(ms)$

To investigate the robustness of this mechanism to the choice of laser parameters, we perform scans over laser peak intensity $I_0$ and pulse duration $\tau_g$. The dependence of the volume-averaged longitudinal field on $I_0$, shown in~\cref{bxav_scan} with asterisk markers, matches well the dependence given by Eq.~(\ref{bx_theory}) and shown with the blue dashed line. The blue dashed line is $| \langle B_x \rangle |$[kT] $= 0.85a_0^2/\sqrt{1 + a_0^2}$. Even at $I_0 \approx 3 \times 10^{19}$W/cm$^2$ ($a_0 = 4$), the axial magnetic field strength can be as high as 5~kT. The pulse duration scan, shown in~\cref{bxav_scan} with square markers, is performed for a fixed peak intensity of $ I_0 \approx 2.1 \times 10^{20}$W/cm$^2$ ($a_0 = 10$). The red dashed line, $| \langle B_x \rangle |$ [kT] $=  0.024\tau_g$[fs], has the same dependence on $\tau_g$ as that given by Eq.~(\ref{bx_theory}). The laser pulse duration is believed to affect the number of hot electrons and, as a result, the magnetic field generation.  For $\tau_g$ as small as 30~fs, we can still get a volume-averaged magnetic field of 1.3~kT. {Additional simulations with a laser wavelength of 1.053~$\micron$ produce similar results, confirming that our scheme is applicable to both Ti:Sa and neodymium-based lasers.}

In summary, we have demonstrated via 3D kinetic simulations a novel mechanism for generating a multi-kT axial magnetic field using multiple regular laser pulses. The twist in the pointing direction of the pulses is the key to driving an azimuthal plasma current that sustains the magnetic field. The twist angle is a convenient control knob for adjusting the direction and magnitude of the axial magnetic field. The field occupies a volume that is tens of thousands of cubic microns and it persists on a ps time scale. The mechanism can be realized for a wide range of laser intensities and pulse durations. Our scheme requires just regular linearly-polarized laser beams, which makes it suitable for implementation {at existing and future multi-kJ PW-class laser facilities that, by design, have to have multiple beamlets~\cite{Kawanaka_2008,Crane_2010,Batani_2014,Morace2019,Yao2022}, including the SG-II UP facility~\cite{zhu2018hpl} that is expected to have multiple kJ-class ps pulses in the near future}.

\bigskip
\section*{\sffamily{Acknowledgements}}
%\begin{acknowledgments}
Y S acknowledges the support by USTC Research Funds of the Double First-Class Initiative, Strategic Priority Research Program of CAS (Grant No. XDA25010200), CAS Project for Young Scientists in Basic Research (Grant No. YSBR060) and Newton International Fellows Alumni follow-on funding. Y S also acknowledge Rui Yan and Robert Kingham for enthusiastic discussions. A. Arefiev's research was supported under the National Science Foundation–Czech Science Foundation partnership by NSF Award No. PHY-2206777. Simulations were performed with EPOCH (developed under UK EPSRC Grants EP/G054950/1, EP/G056803/1, EP/G055165/1, and EP/M022463/1).  The computational center of USTC and Hefei Advanced Computing Center are acknowledged for computational support.
%\end{acknowledgments}
%\nolinenumbers
%\nolinenumbers
%\bibliographystyle{ieeetr}
%\bibliography{aa}



\end{document}